\documentclass[12pt]{article}

\def\myendproof{{\ \vbox{\hrule\hbox{%
   \vrule height1.3ex\hskip0.8ex\vrule}\hrule }}\par}

\newtheorem{theorem}{Theorem}[section]
\newtheorem{lemma}[theorem]{Lemma}
\newtheorem{corollary}[theorem]{Corollary}
\newenvironment{proof}{{\it Proof. }}{\myendproof}

\newcommand{\DD}[0]{{\cal D}}
\newcommand{\RR}[0]{{\cal R}}
\newcommand{\AAA}[0]{{\cal A}}

\newtheorem{claim}{Claim}[section]
\newtheorem{fact}{Fact}

\newcommand{\hp}{\{(h_i,a_i), i \ge 1\}}
\newcommand{\pa}[1]{{\left(#1\right)}}

\newcommand{\ep}[0]{\epsilon}
\newcommand{\la}[0]{\lambda}
\begin{document}

\title{Optimal Constructions of Hybrid Algorithms\thanks{A preliminary
    version of this work appeared in {\em Proceedings of the 5th Annual
      ACM-SIAM Symposium on Discrete Algorithms}, pages 372--381, 1994.
    The first author was supported in part by NSF Grant CCR-9101385.  The
    other three authors were supported in part by AFOSR Contract
    F49620-92-J-0125, DARPA Contract N00014-91-J-1698, and DARPA Contract
    N00014-92-J-1799.  This work was performed while the third and the
    fourth authors were with Department of Mathematics and Laboratory for
    Computer Science, Massachusetts Institute of Technology, Cambridge, MA
    02139.}}

\author{
Ming-Yang Kao\thanks{Department of Computer Science, Yale
  University, New Haven, CT 06520.}
\and 
Yuan Ma\thanks{Haas School of Business, University of California, Berkeley,
  CA 94720.}
\and
Michael Sipser\thanks{Department of Mathematics and Laboratory for Computer
  Science, Massachusetts Institute of Technology, Cambridge, MA 02139.}
\and
Yiqun Yin\thanks{RSA Laboratories, RSA Data Security, 100 Marine World
  Parkway, Redwood Shores, CA 94065.}
}

\date{}

\maketitle

\begin{abstract}
  We study on-line strategies for solving problems with hybrid algorithms.
  There is a problem $Q$ and $w$ {\em basic\/} algorithms for solving $Q$.
  For some $\la \le w$, we have a computer with $\la$ disjoint
  memory areas, each of which can be used to run a basic algorithm and
  store its intermediate results.  In the worst case, only one basic
  algorithm can solve $Q$ in finite time, and all the other basic
  algorithms run forever without solving $Q$.  To solve $Q$ with a
  {\em hybrid\/} algorithm constructed {from} the basic algorithms, we run
  a basic algorithm for some time, then switch to another, and continue
  this process until $Q$ is solved. The goal is to solve $Q$ in the least
  amount of time.  Using {\em competitive ratios\/} to measure the
  efficiency of a hybrid algorithm, we construct an optimal deterministic
  hybrid algorithm and an efficient randomized hybrid algorithm. This
  resolves an open question on searching with multiple robots posed by
  Baeza-Yates, Culberson and Rawlins.  We also prove that our randomized
  algorithm is optimal for $\la=1$, settling a conjecture of Kao, Reif
  and Tate.
\end{abstract}

\section{Introduction} 
We study on-line strategies for solving problems with hybrid
algorithms.  There is a problem $Q$ and $w$ {\it basic} algorithms for
solving $Q$.  For some $\la \le w$, we have a computer with $\la$
disjoint memory areas, each of which can be used to run a basic
algorithm and store its intermediate results.  In the worst case, only
one basic algorithm can solve $Q$ in finite time, and all the other
basic algorithms run forever without solving $Q$.  To solve $Q$ with a
{\em hybrid\/} algorithm constructed from the basic algorithms, we run
a basic algorithm for some time, then switch to another, and continue
this process until $Q$ is solved. The goal is to solve $Q$ in the
least amount of time.

This optimization problem can be conveniently formulated as one of
exploring an unknown environment with multiple robots
\cite{ABM93,BBFY94,BC93,BRS91,FRR90,Pearl84}.  At an origin, there are
$w$ paths leading off into unknown territories. On one of the paths,
there is a goal at an unknown distance $n$ from the origin, and none
of the other paths has a goal.  Initially, there are $\la$ robots
standing at the origin.  The robots can move back and forth on the
paths to search for the goal.  The objective is to minimize the total
distance traveled by all the robots before the goal is found.

We use the notion of a competitive ratio, introduced by Sleator and
Tarjan~{\cite{ST85}}, to measure the efficiency of an exploration algorithm
$\AAA$.  Let cost($\AAA$) be the (worst-case or expected) total distance
traveled by all the robots using $\AAA$.  Given a constant $c$, we say that
$\AAA$ has a {\it competitive ratio} $c$ if $\mbox{cost}(\AAA) \le
c{\cdot}n + o(n)$.

An extreme case where there is only one robot, i.e., $\la=1$, was
studied by many researchers.  In particular, Baeza-Yates, Culberson and
Rawlins~\cite{BCR93} presented an optimal deterministic algorithm.  Kao,
Reif and Tate~\cite{KaoRT92.iof} reported a randomized algorithm, proved
its optimality for $w=2$, and conjectured its optimality for all $w >2$.
In these two algorithms, the single robot searches the $w$ paths in a
cyclic fashion, and the returning positions on the paths form a geometric
sequence. (A returning position on a path refers to a point where the robot
starts moving towards the origin after advancing away from it.)

Another extreme case where the number of robots equals the number of paths,
i.e., $\la = w$, was studied by Azar, Broder and Manasse~\cite{ABM93}, who
showed that the smallest competitive ratios are $w$ for both deterministic
and randomized algorithms.  This ratio $w$ can be achieved by the simple
algorithm in which each robot explores a single path and keeps moving
forward until the goal is found.

The general case $1 < \la < w$ has not been well understood, and its
difficulty is twofold.  Since $\la > 1$, an exploration algorithm must
coordinate its robots in order to achieve optimality. Moreover, since $\la
< w$, some robots must move back and forth on two or more paths, and their
returning positions on those paths are crucial for optimality.  Our main
results are that for all values of $\la$ and $w$, we construct
\begin{itemize}
\item a deterministic algorithm with the smallest possible competitive
  ratio, and
\item an efficient randomized algorithm that is provably optimal for $\la =
  1$.
\end{itemize}
Our deterministic algorithm resolves an open question of Baeza-Yates et
al.~\cite{BCR93} on exploration using multiple robots.  The optimality
proof for our randomized algorithm with $\la=1$ settles the conjecture of
Kao et al.~\cite{KaoRT92.iof} in the affirmative.  Our results also imply
that randomization can help reduce the competitive ratios if and only if $
\la < w$.

We discuss our deterministic exploration algorithm in
\S\ref{sec_deterministic} and the randomized algorithm in
\S\ref{sec_randomized}.  The paper concludes with some directions for
future research in \S\ref{sec_conclusion}.

Throughout the paper, we label the $w$ paths by $0, 1, \ldots, w-1$ and the
$\la$ robots by $1, 2, \ldots \la$.

\section{An optimal deterministic exploration algorithm}
\label{sec_deterministic}
Let $D(w,\la)$ denote the smallest competitive ratio for all deterministic
exploration algorithms.  The main result of this section is the next
theorem.
\begin{theorem}  \label{thm-deterministic}
\(D(w,\la) = \la + 2 {\left(w-\la+1\right)^{w - \la+1} \over \left(w-\la\right)^{w-\la}}.\)
\end{theorem}

Baeza-Yates et al.~\cite{BCR93} studied the case $\la = 1$, and their
results can be restated as 
\begin{eqnarray}\label{eq-BCR-original} D(w,1) = 1 + 2
  \frac{w^w}{\left(w-1\right)^{w-1}}.
\end{eqnarray}
Theorem~\ref{thm-deterministic} generalizes (\ref{eq-BCR-original})
and answers the open question in~{\cite{BCR93}} on optimal exploration
using multiple robots.  To prove Theorem~\ref{thm-deterministic}, we
describe our deterministic algorithm in \S\ref{subsec-det-alg} and give a
lower bound proof in \S\ref{subsec-det-lower}.

\subsection{A deterministic exploration algorithm} \label{subsec-det-alg}
We first review the exploration algorithm for $\la= 1$ given by
Baeza-Yates et al.~{\cite{BCR93}}.  This algorithm, referred to as
$\DD(w,1)$, is used as a subroutine in our algorithm for general $\la$.
Let
\[
f(w, i) = 
\left\{
\begin{array}{cl}
  \left( \frac{w}{w-1} \right)^i & \mbox{for} \ i \ge 0, \\ 0 & \mbox{for}
  \ i < 0.
\end{array}
\right.
\]
In algorithm $\DD(w,1)$, the single robot searches the $w$ paths in a fixed
cyclic order. The search proceeds in $stages$, starting from stage 0.  In
stage $i$, the robot searches path $i \bmod w$ until position $f(w,i)$ and
moves back to the origin if the goal is not found by then. If the goal is
at position $f(w,i)+1$, then the robot finds it in stage $i+w$, traveling a
total distance of $f(w,i)+1 + 2 \sum_{j=0}^{i+w-1} f(w,j)$.  Baeza-Yates et
al.\ showed that the smallest competitive ratio of $\DD(w,1)$ is
\begin{eqnarray}
  \overline{\lim}_{i \rightarrow \infty} \frac{f(w,i)+1 + 2 
    \sum_{j=0}^{i+w-1} f(w,j)}{f(w,i)+1} \le 1 + 2 {w^w \over
    \left(w-1\right)^{w-1}}.
\label{eq-BCR-general} 
\end{eqnarray}

In our algorithm for general $\la$, for each $k < \la$, the $k$-th robot
only searches path $k$. These $\la-1$ robots simply advance on their own
paths and never move towards the origin.  Let $ w'=w-\la+1$.  The $\la$-th
robot explores the remaining $w'$ paths using $\DD(w',1)$.  The
algorithm proceeds in rounds until the goal is found.  In the $i$-th round,
the robots move as follows:
\begin{itemize}
\item The $\la$-th robot chooses some path $p$ according to $\DD(w',1)$
  and searches it from the origin to position $f(w',i-w')$.
\item All the robots then move in parallel from position $f(w',i-w')$ to
  position $f(w',i+1-w')$ on the paths where they stand.
\item The $\la$-th robot continues to search path $p$ from position
  $f(w',i+1-w')$ to position $f(w',i)$ and then moves back to the origin.
\end{itemize}

We next analyze the above algorithm. If the goal is at position
$f(w',i-w')+1$ on some path, it is found in round $i$. By the time the goal
is found, the first $\la-1$ robots have each traveled a distance of
$f(w',i-w')+1$, and the $\la$-th robot a distance of $f(w',i-w')+1 +
2 \sum_{j=0}^{i-1} f(w',j)$.  Hence, the smallest competitive ratio
of the above exploration algorithm is
\[\overline{\lim}_{i \rightarrow \infty}
\frac{ \la {\cdot} \pa{f(w',i-w')+1} + 2 \sum_{j=0}^{i-1} 
f(w',j)}{f(w',i-w')+1}.
\]
By (\ref{eq-BCR-general}), the above formula is upper bounded
by
\[ \la + 2 {{w'}^{w'} \over \left(w'-1\right)^{w'-1}}
= \la + 2 
{\left(w-\la+1\right)^{w - \la+1} \over \left(w-\la\right)^{w-\la}},
\]
which  equals  the competitive ratio stated in
Theorem~\ref{thm-deterministic}.

\subsection{A matching lower bound}
\label{subsec-det-lower} 
Here, we prove that our deterministic algorithm in \S\ref{subsec-det-alg}
is optimal by deriving a matching lower bound on the smallest competitive
ratio $r_\AAA$ of any arbitrary deterministic exploration algorithm
$\AAA(w,\la)$ with $w$ paths and $\la$ robots.  Let $t_0$ to be the
time when $\AAA(w,\la)$ commences.  For $i > 0$, $t_i$ denotes the
$i$-th time when a robot of $\AAA(w,\la)$ starts moving towards the
origin on some path.  $\AAA(w,\la)$ can be partitioned into {\em
  phases\/} where phase $i$ starts at $t_{i-1}$ and ends at $t_i$.  The
notion of a phase differs {from} that of a round in \S\ref{subsec-det-alg}.

\begin{lemma} \label{lem-preliminary}
  Given $\AAA(w,\la)$, there is a deterministic exploration algorithm
  ${\cal A}'\left(w,\la\right)$ such that $r_{{\cal A}'} \le r_{\cal A}$ and ${\cal
    A}'\left(w,\la\right)$ satisfies the following three properties:
\begin{itemize}
\item No two robots search the same path in the same phase.
\item No robot moves towards the origin if some robot stays at the origin.
\item As soon as a robot starts moving towards the origin on some path, all
  the other robots stop moving until that robot moves back to the origin
  and then advances on another path to a previously unsearched location.
\end{itemize}
\end{lemma}
\begin{proof}
  Straightforward.
\end{proof}
If $\la =1$, i.e., there is only one robot, then $\AAA(w,1)$ can be
characterized by a sequence $\hp$ where $a_i$ is the index of the path
where the robot starts moving towards the origin in phase $i$, and $h_i$ is
the distance that the robot searches on path $a_i$ in phase $i$.  By simple
calculation, the ratio $r_{\cal A}$ of $\AAA(w,1)$ equals
\begin{eqnarray}
\label{eq-motivation} 1 + 2{\cdot}\overline{\lim}_{i \rightarrow
\infty} \left( \frac{h_1+ \cdots + h_{i'-1}}{h_i} \right),
\end{eqnarray}
where $i'$ is the smallest index such that $i' > i$ and $a_{i'} = a_i$.
Motivated by this finding, for any given $\hp$, we define the corresponding
{\em ratio sequence\/} $\{H_i, i \ge 1\}$ by
\begin{eqnarray}\label{eq-H-def} H_i
= \frac{h_1+ \cdots + h_{i'-1}}{h_i} 
\end{eqnarray}
where $i'$ is the smallest index with $i' > i$ and $a_{i'} = a_i$.  Using
$H_i$, (\ref{eq-motivation}) can be written as
\[
1 + 2{\cdot}\overline{\lim}_{i \rightarrow \infty} H_i.
\]

A sequence $\hp$ is a {\em $w$-sequence\/} if $h_i > 0$ and $a_i$ is an
integer for all $i$ as well as $|\{i \mid a_{i} = j\}| = \infty$ for at
least $w$ integers $j$.  A $w$-sequence $\hp$ is a {\em cyclic sequence\/}
if $a_i = i \bmod w$.  Since the integers $a_i$ are uniquely specified in a
cyclic sequence, we represent a cyclic sequence by $\{s_i, i \ge 1\}$.  The
corresponding {\em ratio sequence}, denoted by $\{S_i, i \ge 1\}$, is thus
defined by
\begin{eqnarray}
\label{eq-S-def} S_i = \frac{s_1+ \cdots + s_{i+w-1}}{s_i}.
\end{eqnarray}

\begin{fact} $($see {\rm \cite{BCR93}}$)$ \label{fact_BCR}
  For every cyclic $w$-sequence, $\overline{\lim}_{i \rightarrow \infty}
  S_i \ge \frac{w^w}{\left(w-1\right)^{w-1}}.$
\end{fact}

\begin{lemma} \label{lem-sequence}
  For each $w$-sequence $\hp$, there exists a cyclic $w$-sequence $\{s_i, i
  \ge 1 \}$ such that $\overline{\lim}_{i \rightarrow \infty} H_i \ge
  \overline{\lim}_{i \rightarrow \infty} S_i$.
\end{lemma}
\begin{proof}
  See Appendix~\ref{app-sequence}.
\end{proof}

Intuitively, Lemma~\ref{lem-sequence} shows that we can modify a
deterministic exploration algorithm to search the paths in a cyclic order
without increasing its competitive ratio.

\begin{lemma} \label{lem-algorithm-sequence}
  If $\AAA(w,\la)$ has a finite competitive ratio and satisfies the
  properties of Lemma~\ref{lem-preliminary}, then there exists a
  $\left(w-\la+1\right)$-sequence $\hp$ such that
\[r_{\cal A} \ge
\la + 2{\cdot}\overline{\lim}_{i \rightarrow \infty} H_i,\] where $\{H_i, i
\ge 1\}$ is as defined in {\rm (\ref{eq-H-def})}.
\end{lemma}
\begin{proof}
  First, we inductively define the sequence $\hp$ and a sequence of
  $w$-dimensional vectors $\pi_{i}=(\pi_{i}(0),\ldots,\pi_{i}(w-1))$.
  Then, we prove the inequality claimed in the lemma.

We first define $(h_1,a_1)$ and $\pi_1$ by looking at the first phase of
$\AAA(w,\la)$. Assume that at time $t_1$, a robot starts moving
towards the origin on path $j$.  Let $h_1$ be the distance that the robot
has searched on path $j$.  Define
\[
\pi_1 = (0,1,2,\ldots,w-1) \ \mbox{and} \ a_1 = \pi_{1}(j).
\]
Once $\pi_{i-1}$ is defined, we define $\left(h_i, a_i\right)$ and
$\pi_{i}$ by looking at phase $i$ of $\AAA(w,\la)$. Recall that phase $i$
starts at time $t_{i-1}$ and ends at time $t_{i}$.  Assume that path $l$ is
the unique path that is not searched in phase $i-1$ but is searched in
phase $i$.  Also assume that at time $t_i$, a robot starts moving towards
the origin on path $k$.  Let $h_i$ be the distance that the robot has
searched on path $k$. For $j = 0, \ldots, w-1$, let
\[
\pi_{i}(j) = \left\{
\begin{array}{ll}
\pi_{i-1}(l) & \mbox{if} \ j = k, \\
\pi_{i-1}(k) & \mbox{if} \ j = l, \\
\pi_{i-1}(j) & \mbox{if} \ j \ne k \ \mbox{and} \ j \ne l;
\end{array}
\right.
\]
i.e., we switch the $k$-th entry and the $l$-th entry of $\pi_{i-1}$ to
obtain $\pi_{i}$. Let $a_i = \pi_i(k)$.

Since $\AAA(w,\la)$ has a finite competitive ratio, $\hp$ is an infinite
sequence. For every $i \ge 1$, $t_i$ is the time when a robot starts moving
towards the origin on some path $p$.  Let $i' > i$ be the index such that
phase $i'$ is the first phase that path $p$ is searched again after $t_i$.
Such $i'$ exists and is finite.

\begin{claim} \label{claim-i'}
For all $i \ge 1$, $a_{i'} = a_{i}$ and $a_{j} \ne
a_{i}$ for $j = i+1, \ldots, i'-1$.
\end{claim}
To prove this claim, assume that $a_i = \pi_i(k)$, i.e., a robot
starts 
moving towards the origin on path $k$ immediately after $t_i$. In phase
$i+1$, that robot moves back on path $k$ to the origin and then searches
another path $l$ with $l \not = k$.  Then, some robot starts moving towards
the origin on path $k_1$ with $k_1 \not = k$ right after $t_{i+1}$. Since
$k\not = l$ and $k\not = k_1$, by the definition of $\pi_{i+1}$,
\[
\pi_{i+1}(k) = \pi_{i}(k).
\]
By the choice of $i'$, path $k$ must be idle from $t_{i+1}$ to $t_{i'-1}$.
Hence,
\begin{eqnarray}\label{eq-check-1} 
  \pi_{j}(k) = \pi_{j-1}(k) = \cdots = \pi_{i+1}(k) =
  \pi_{i}(k) = a_i.
\end{eqnarray}
Moreover, $a_j \ne \pi_{j}(k)$.  Thus, from (\ref{eq-check-1}),
\begin{eqnarray}\label{eq-non-equal}
  a_j \ne a_i \ \mbox{for} \ j = i+1, \ldots, i'-1.
\end{eqnarray}
By the choice of $i'$, path $k$ is reused in phase $i'$. Assume that a
robot searches path $k_2$ immediately after $t_{i'}$. By the inductive
procedure for defining $\pi_{i'}$ and $a_{i'}$,
\begin{eqnarray}
\label{eq-check-3} 
\pi_{i'}(k_2) = \pi_{i'-1}(k) \ \mbox{and} \ a_{i'} = \pi_{i'}(k_2).  
\end{eqnarray}
Combining (\ref{eq-check-1}) and (\ref{eq-check-3}), we have
$a_{i'}=a_i$.  This and (\ref{eq-non-equal}) conclude the proof of
Claim~\ref{claim-i'}.

\begin{claim} \label{claim-w-sequence}
$\hp$ is a $\left(w-\la+1\right)$-sequence.
\end{claim}
To prove this claim, the only nontrivial property of the sequence that we
need to verify is that there are at least $\left(w-\la+1\right)$ integers $j$ such
that $ | \{i \mid  a_i = j \} | =+\infty$. By Claim~\ref{claim-i'}, it
suffices to prove that there exist $\left(w-\la +1\right)$ integers $j$ such that
\begin{equation} \label{eq-goal-next}
  a_i = j \ \mbox{for some} \ i.
\end{equation}
Without loss of generality, we label the $w$ paths in such a way that 
\begin{itemize}
\item the label of the path where a robot starts moving towards the origin
  immediately after $t_1$ is 0;
\item paths $1, 2, \ldots, w-\la$ are not searched before $t_1$;
\item $i_1 < i_2 < \cdots < i_{w-\la}$ where $i_j$ is the first phase in
  which path $j$ is searched.
\end{itemize}
By the assumption on $a_0$ and the definition of $a_1$,
\begin{equation} \label{eq-a-1}
a_1 = 0.
\end{equation}
For $j = 1, \ldots, w-\la$, let $j^*$ be the label of the path where a
robot starts moving towards the origin immediately after $t_{i_j}$. By the
definitions of $\{\pi_i, \ i \ge 1 \}$ and $\hp$,
$
\pi_{i_j}(j^*) = \pi_{i_{j}-1}(j) = \cdots = \pi_1(j) = j.
$
Therefore, 
\[
  a_{i_j} = \pi_{i_j}(j^*) = j \ \mbox{for} \ 1 \le j \le w - \la.
\]
By this and (\ref{eq-a-1}), at least $w-\la+1$ integers $j$ satisfy
(\ref{eq-goal-next}), finishing the proof of
Claim~\ref{claim-w-sequence}.

Continuing the proof of Lemma~\ref{lem-algorithm-sequence}, for each $i \ge
1$, let $p$ be the path where a robot starts moving towards the origin
right after $t_i$. Let $T$ be the first time when path $p$ is searched for
exactly distance $h_i$ in phase $i'$.  By the properties stated in
Lemma~\ref{lem-preliminary}, the $\la$ robots stand at different paths at
time $T$. Let $d_1, d_2, \ldots, d_{\la-1}$ be the distances that the
robots except the one on path $p$ are from the origin at time $T$.  Let
$d_j = \min \{d_1, d_2, \ldots, d_{\la-1}\}$.  There are two cases.

{\it Case 1}: $d_j \ge h_i$. If the goal is on path $p$ at distance
$h_i+1$, then when $\AAA(w,\la)$ finds the goal, its performance ratio
is at least
\begin{eqnarray}
& & \frac
{d_1 + d_2 + \cdots + d_{\la-1}}{h_i +1} + \frac{2 
\left(h_1 + \cdots + h_{i'-1}\right) + h_i + 1} {h_i + 1} \nonumber \\
& \ge & {h_i \over h_i + 1} \left(\la + 
{ 2 \left(h_1 + {\cdots} + h_{i'-1}\right) \over h_i} \right). 
  \label{eq-case-1}
\end{eqnarray}

{\it Case 2}: $d_j < h_i$. Let $p'$ be a path that has been searched up to
distance $d_j$ at time $T$.  If the goal is on path $p'$ at distance
$d_j+1$, then when $\AAA(w,\la)$ finds the goal, its performance ratio is
at least
\begin{eqnarray} 
& & \frac
{d_1 + d_2 + \cdots + d_j + 1 + \cdots + d_{\la-1}}{d_j +1} 
 + \frac{2 \left(h_1 + \cdots + h_{i'-1}\right) + h_i} {d_j + 1}
 \nonumber \\
& \ge & {d_{j} +1 \over d_{j}} 
\left(\la + \frac{ 2 \left(h_1 + {\cdots} + h_{i'-1}\right)}{h_i} \right).
 \label{eq-case-2}  
\end{eqnarray}
Since $\AAA(w,\la)$ has a finite competitive ratio,
\(
\lim_{i \rightarrow \infty} h_i = \lim_{j \rightarrow \infty} d_j = +
  \infty.
\)
Therefore, by (\ref{eq-case-1}) and (\ref{eq-case-2}),
\[
r_{\cal A} \ge \overline{\lim}_{i \rightarrow \infty} \left( \la + \frac{2
    \left(h_1 + \cdots + h_{i'-1}\right)}{h_i} \right) = \la +
2{\cdot}\overline{\lim}_{i \rightarrow \infty} H_i. 
\]
This and Claims~\ref{claim-i'} and~\ref{claim-w-sequence} conclude the
proof of Lemma~\ref{lem-algorithm-sequence}
\end{proof}

Fact~\ref{fact_BCR} and Lemmas \ref{lem-preliminary}, \ref{lem-sequence}
and \ref{lem-algorithm-sequence} give a lower bound proof for
Theorem~\ref{thm-deterministic}.

\section{A randomized exploration algorithm} \label{sec_randomized} 
We give our randomized exploration algorithm for general $\la$ in
\S\ref{subsec_alg_general} and prove its optimality for $\la =1$ in
\S\ref{subsec_k_1}.

\subsection{A randomized exploration algorithm for general $\la$}
\label{subsec_alg_general}
We first review the randomized search algorithm of Kao et
al.~\cite{KaoRT92.iof} for $\la = 1$, which we refer to as $\RR(w,1)$.
Choose $r_{w} > 1$ such that
\[
\frac{r_{w}^w-1}{\left(r_{w}-1\right)\ln r_{w}} = 
\min_{r > 1}\frac{r^w-1}{\left(r-1\right)\ln r}.
\]
Such $r_w$ exists and is unique \cite{KaoRT92.iof}.  $\RR(w,1)$
proceeds as follows:
\begin{enumerate}
\item $\sigma \! \leftarrow \!$ a random permutation of $\{0, \ldots ,w-1\}$;
\item $\ep \! \leftarrow \!$ a random real number uniformly chosen
  {from} $[0,1)$;
\item $d\leftarrow r_{w}^\ep$;
\item $i \leftarrow 1$;
\item repeat
\\ \hspace*{0.3in}  explore path $\sigma\left(i\right)$ up to distance $d$;
\\ \hspace*{0.3in} if goal not found then return to origin;
\\ \hspace*{0.3in} $d\leftarrow d{\cdot}r_w$;
\\ \hspace*{0.3in} $i\leftarrow \left(i+1\right)\bmod w$;
\item[] until the goal is found.
\end{enumerate}
Let $R\left(w,\la\right)$ denote the smallest competitive ratio for randomized
exploration algorithms for $w$ paths and $\la$ robots.  Let
\[ 
\bar{R}(w) = 1 + \frac{2}{w}{\cdot}\frac{r_w^w-1}{\left(r_w-1\right)\ln r_w}.
\]
\begin{fact} $($see {\rm \cite{KaoRT92.iof}}$)$ \label{fact_KRT}
  $R\left(w,1\right) \leq \bar{R}(w).$
\end{fact}

We now construct our randomized exploration algorithm for general $\la$
using $\RR(w,1)$ as a subroutine.  First, we pick a random permutation
$\sigma$ of $\{0,1,\ldots,w-1\}$. For the first $\la-1$ robots, robot $i$
searches only path $\sigma\left(i\right)$.  These robots search their own paths at the
same constant speed.  The $\la$-th robot searches the remaining $w-\la+1$
paths using $\RR(w-\la+1,1)$ also at a constant speed. The speeds are
coordinated by a parameter $v$ such that at any given time, the total
distance traveled by the $\la$-th robot is $v$ times the distance traveled
by each of the first $\la-1$ robots.  By choosing an appropriate $v$, we
can prove the next theorem.

\begin{theorem} \label{thm_general_R}
  $R\left(w,\la\right) \leq \frac{1}{w}\pa{ \left(\la-1\right) + \sqrt{\left(w-\la+1\right) \ 
      \bar{R}(w-\la+1)}}^2$.
\end{theorem}
\begin{proof}
  The smallest competitive ratio of our exploration algorithm is at most
\[
{\la -1 \over w} \left( \left(\la -1\right) + v \right) + {w-\la+1 \over w}
\left( {\la -1 \over v} + 1 \right) \bar{R}(w-\la+1).
\]
At $v = \sqrt{\left(w - \la +1\right) \bar{R}(w-\la+1)}$, this expression assumes
its minimum
\[
\frac{1}{w}\pa{ \left(\la-1\right) + \sqrt{\left(w-\la+1\right) \ 
    \bar{R}(w-\la+1)}}^2.
\]
\end{proof}

Combining Theorem~\ref{thm-deterministic} and~\ref{thm_general_R}, we can
prove the following corollary.

\begin{corollary}
  If $\la < w$, then the smallest competitive ratio of randomized
  exploration algorithms is always smaller than that of deterministic ones.
\end{corollary}
\begin{proof}
  By Theorems~\ref{thm-deterministic} and~\ref{thm_general_R}, we only need
  to prove
\begin{eqnarray}\label{inequ-corollary}
\frac{1}{w}\pa{ \left(\la-1\right) +
\sqrt{\left(w-\la+1\right) \bar{R}(w-\la+1)}}^2 
< \la + 2 \ {\left(w-\la+1\right)^{w - \la+1} \over \left(w-\la\right)^{w-\la}}.
\end{eqnarray}
Let $a = \bar{R}(w-\la+1)$, by Fact~\ref{fact_KRT}, which is a
competitive ratio of a randomized exploration algorithm with one robot and
$w-\la+1$ paths.  Let
$b=1+2{\left(w-\la+1\right)^{w-\la+1}\over\left(w-\la\right)^{w-\la}}$, by
Theorem~\ref{thm-deterministic}, which is the smallest competitive ratio of
deterministic algorithms also for one robot and $w-\la+1$ paths.  Since $a
< b$ \cite{KaoRT92.iof}, to prove (\ref{inequ-corollary}), it is
suffice to show that
\begin{eqnarray*} 
\frac{1}{w}\pa{ \left(\la-1\right) +
\sqrt{\left(w-\la+1\right) \ a}}^2 
\le \la -1 + a,
\end{eqnarray*}
which is equivalent to $\left(a - \sqrt{w-\la+1}\right)^2 \ge 0$.
\end{proof}

\subsection{A matching lower bound for $\la = 1$} \label{subsec_k_1} 
The next theorem shows that our randomized algorithm is optimal for
$\la=1$.

\begin{theorem} \label{thm_difficult}
  $R\left(w,1\right) \geq \bar{R}(w).$
\end{theorem}

The proof of this theorem uses a strengthened version of
Fact~\ref{fact_KRT_lb} below.  Let $\vec{s}=\{s_i, i \geq 0\}$ denote an
infinite sequence of positive numbers.  Let
\[
S_w = \{ \{s_i, i \geq 0\} | \lim_{i\rightarrow \infty} s_i = \infty,
s_0=1, \ \mbox{and for all} \ i \geq 0, s_{i+w} > s_i\}.
\]  
For any $\ep > 0$ and $\vec{s} = \{s_i, i \geq 0\} \in S_w$, let
\[
G_w(\ep,\vec{s}) = \ep 
\sum_{i=0}^{\infty}\frac{s_i+{\cdots}+s_{i+w-1}}{s_i^{1+\ep}}.
\]

An exploration algorithm with one robot is called {\it cyclic} if it
searches the paths one after another in a cyclic order.

\begin{fact} \label{fact_KRT_lb} $($see {\rm \cite{KaoRT92.iof}}$)$
  The smallest competitive ratio of any cyclic randomized exploration
  algorithm is at least
\[
\sup_{\ep > 0} \inf_{\vec{s} \in S_w}\{1 + \frac{2}{w} \ 
G_w(\ep,\vec{s})\}.
\]
\end{fact}

Note that if $w=2$, i.e., there are only two paths, then every exploration
algorithm is cyclic. This is not true for $w \geq 3$.  Ma and
Yin~\cite{MY94} strengthened Fact~\ref{fact_KRT_lb} by removing the cyclic
assumption about exploration algorithms.

\begin{fact} \label{fact_MY} $($see {\rm \cite{MY94}}$)$
  The lower bound stated in Fact {\rm \ref{fact_KRT_lb}} also holds for any
  arbitrary randomized exploration algorithms that may or may not be
  cyclic.
\end{fact}
\begin{proof}
  By Yao's formulation of von Neumann's minimax principle~\cite{Yao77}, it
  suffice to lower bound the competitive ratios of all deterministic
  algorithms against a chosen probability distribution.  The proof idea is
  to show that an optimal algorithm against the distribution used by Kao et
  al.~\cite{KaoRT92.iof} for Fact~\ref{fact_KRT_lb} can be modified to be
  cyclic.
\end{proof}

By Fact~\ref{fact_MY}, to prove Theorem~\ref{thm_difficult}, we only need
to prove the next theorem. Let
\[
C_w = \frac{r_w^w-1}{\left(r_w-1\right)\ln r_w}.
\]
\begin{theorem}\label{thm_main}
  $\sup_{\ep>0}\inf_{\vec{s}\in S_w} G_w(\ep,\vec{s}) \geq C_w$.
\end{theorem}

The proof of Theorem~\ref{thm_main} is divided into three parts.  In
\S\ref{subsub_1}, we lower bound the infinite sum $G_w(\ep,\vec{s})$ by the
finite sum
\[
H(k,\vec{s}(\ep)) = \frac{-\ep+
  \sum_{i=0}^{k-1}\frac{s_i(\ep)+{\cdots}+s_{i+w-1}(\ep)}{\left(s_i(\ep)\right)^{1+\ep}}}
{\ln s_k(\ep)}.
\] 
In \S\ref{subsub_2}, we lower bound this finite sum by $C_w$. Finally in
\S\ref{subsub_3}, we complete the proof of Theorem~\ref{thm_main}.

\subsubsection{Lower bounding $G_w(\ep,\vec{s})$ by
  $H(k,\vec{s}(\ep))$}\label{subsub_1}\ 

\begin{lemma}\label{lem_step1}
  For all $\ep > 0$, there exists $\{s_i(\ep),i \geq 0\} \in S_w$ such that
  for all $k$ with $s_k(\ep) > 1$,
\[
\inf_{\vec{s} \in S_w} G_w(\ep,\vec{s}) \geq H(k,\vec{s}(\ep)).
\]
\end{lemma}
\begin{proof}
  By the definition of infimum, for all $\ep >0$, there exists
  $\vec{s}(\ep) = \{s_i(\ep), i \geq 0\} \in S_w$ such that
\begin{eqnarray*}
& & \inf_{\vec{s}\in S_w} G_w(\ep,\vec{s}) + \ep^2 \\
& \geq & G_w(\ep,\vec{s}(\ep))
\\
& = & \ep \sum_{i=0}^{k-1}
\frac{s_i(\ep)+{\cdots}+s_{i+w-1}(\ep)}{\left(s_i(\ep)\right)^{1+\ep}}
 + \left(s_k(\ep)\right)^{-\ep} \left(\ep
\sum_{i=k}^\infty\frac{\frac{s_i(\ep)}{s_k(\ep)}+{\cdots}
+\frac{s_{i+w-1}(\ep)}{s_k(\ep)}}
{\left(\frac{s_i(\ep)}{s_k(\ep)}\right)^{1+\ep}}\right).
\end{eqnarray*}
Since $\vec{s'}(\ep) = \{s'_i(\ep) = \frac{s_{k+i}(\ep)}{s_k(\ep)}, i \geq
0\}$ is in $S_w$,
\begin{eqnarray*}
\inf_{\vec{s} \in S_w} G_w(\ep,\vec{s}) + \ep^2 
& = & 
\left(\ep
\sum_{i=0}^{k-1}
\frac{s_i(\ep)+\cdots+s_{i+w-1(\ep)}}{\left(s_i(\ep)\right)^{1+\ep}}\right) 
 + \left(s_k(\ep)\right)^{-\ep} G_w(\ep,\vec{s'})
\\
& \geq & \left(\ep
\sum_{i=0}^{k-1}\frac{s_i(\ep)+\cdots+s_{i+w-1}(\ep)}
{\left(s_i(\ep)\right)^{1+\ep}}\right) + \left(s_k(\ep)\right)^{-\ep} \inf_{\vec{s}\in S_w} G_w(\ep,\vec{s}).
\end{eqnarray*}
Since $s_k(\ep)>1$ and $\frac{\ep}{1-x^{-\ep}} \geq \frac{1}{\ln x}$ for
all $x>1$,
\begin{eqnarray*}
  \inf_{\vec{s} \in S_w} G_w(\ep,\vec{s}) & \geq &
  \frac{\ep}{1-\left(s_k(\ep)\right)^{-\ep}} \left(-\ep+\sum_{i=0}^{k-1}
    \frac{s_i(\ep)+\cdots+s_{i+w-1}(\ep)}
    {\left(s_k(\ep)\right)^{1+\ep}}\right)
\\ & \geq & \frac{1}{\ln s_k(\ep)}\left(-\ep+\sum_{i=0}^{k-1}
  \frac{s_i(\ep)+\cdots+s_{i+w-1}(\ep)}{\left(s_k(\ep)\right)^{1+\ep}}\right).
\end{eqnarray*}
\end{proof}

\begin{lemma}\label{lem_an}
  There exist a strictly increasing integer sequence $\{p_i, i \geq 0\}$
  and a sequence $\vec{s}(\ep_n)$ such that
\[
\lim_{n\rightarrow\infty} H(k_n,\vec{s}(\ep_n)) \ \mbox{exists and is
  finite}
\]
and
\[
\sup_{\ep>0}\inf_{\vec{s}\in S_{w}} G_w(\ep,\vec{s}) \geq
\lim_{n\rightarrow\infty}H(k_n,\vec{s}(\ep_n)),
\]
where \(\ep_n= {1 \over {p_n^2}}\), \(0 < p_n \leq k_n \leq p_n+w -1,\) and
\(s_{k_n}(\ep_n) = \max \{s_{p_n}(\ep_n), \ldots, s_{p_n+w-1}(\ep_n)\}.\)
\end{lemma}
\begin{proof}
  This lemma follows from the fact that by Fact~\ref{fact_MY} and
  Lemma~\ref{lem_step1}, $H(k,\vec{s}(\ep))$ is bounded.
\end{proof}

\subsubsection{Lower Bounding $H(k,\vec{s}(\ep))$ by $C_w$}
\label{subsub_2}

This is the most difficult part of the proof of Theorem~\ref{thm_main} and
requires six technical lemmas, namely, Lemmas \ref{lem_geo} through
\ref{lem_L_rest}. The proofs of these lemmas are given in
Appendix~\ref{app-main}.

We first rewrite
\begin{eqnarray*}
H(k_n,\vec{s}(\ep_n)) & = & {{1} \over {\ln
      s_k(\ep_n)}} 
 \left( -\ep_n  + \sum_{i=0}^{k_n-1}{{{s_i(\ep_n)+{\cdots}+s_{i+w-1}(\ep_n) }}
 \over {\left(s_i(\ep_n)\right)^{1+\ep_n}}} \! \right) \\
 & = &  {{1} \over {\ln s_k(\ep_n)}}
 \left( -\ep_n + \sum_{j=0}^{w-1} \sum_{i=0}^{k_n-1} 
{{s_{i+j}(\ep_n)} \over {\left(s_i(\ep_n)\right)^{1+\ep_n}}} \right).
\end{eqnarray*}
Hence, by Lemma \ref{lem_an}, there is a constant $C$ such that
\begin{equation}\label{equ_break_H}
C \geq H(k_n,\vec{s}(\ep_n)) \geq {{-\ep_n+\sum_{j=0}^{w-1}L_n(j)} 
\over {\ln s_k(\ep_n)}},
\end{equation}
where 
\[
L_n\left(0\right) = \sum_{i=0}^{k_n-1} {1 \over
{\left(s_i(\ep_n)\right)^{\ep_n}}}
\]
and \[ 
L_n(j) =
\sum_{i=0}^{k_n-j}{{s_{i+j}(\ep_n)}\over{\left(s_i(\ep_n)\right)^{1+\ep_n}}}\
\]
for $j = 1, \ldots, w-1$.  Note that $s_{k_n} > 1$ implies $\ln s_{k_n} >
0$.  

To further lower bound $H(k_n, \vec{s}(\ep_n))$, we first work on
$L_{n}(0)$ to show that $s_{k_n}(\ep_n) \rightarrow \infty$ as $n
\rightarrow \infty$.  Then, we work on $L_{n}(1)$ to show that
$s_{n}(\ep_n)$ grows somewhat smoothly at a rate exponential in $n$.
Finally, we lower bound all $L_{n}(j)$.

The next lemma is frequently used in this section.

\begin{lemma}\label{lem_geo}
For every positive integer $m$ and for all $\ep,
x_0,\ldots,x_{m}>0$,
\[
{x_1 \over x_0^{1+\ep}} +
{x_2 \over x_1^{1+\ep}} +{\cdots}+
{x_m \over x_{m-1}^{1+\ep}}
\ge 
{m \over \left(1+\ep\right)^m} \left({x_m^{\left({1 \over {1+\ep}}\right)^m} \over
x_0}\right)^{E_\ep\left(m\right)} \]
where $ 
E_\ep\left(m\right)={\ep\left(1+\ep\right)^m \over {\left(1+\ep\right)^m-1}}.$
\end{lemma}

The next two lemmas give some properties of the sequence 
$\{s_{k_n}(\ep_n), k \geq 0\}$.

\begin{lemma}\label{lem_s_k_big}
\(\lim_{n\rightarrow\infty} s_{k_n}(\ep_n) = \infty.\)
\end{lemma}

For all $n$, pick $h_n \in \{k_{n}-w+1,\ldots,k_n-1\}$ such that
\[
s_{h_n}(\ep_n) = \min\{s_{k_n-w+1}(\ep_n),\ldots,s_{k_n-1}(\ep_n)\}.
\] 
Since $p_n \rightarrow \infty$ as $n \rightarrow \infty$, we assume without
loss of generality that $k_n-w+1 \geq 0$.  Since $s_{k_n}(\ep_n) > 1$,
there exists a unique $\nu_n$ such that
\[
s_{h_n}(\ep_n) = \left(s_{k_n}(\ep_n)\right)^{1-\nu_n}.
\] 
By the choice of $k_n$ and the monotonicity of $S_w$, $s_{k_n}(\ep_n) \geq
s_{h_n}(\ep_n)$ and thus $\nu_n \geq 0$.

\begin{lemma} \label{lem_nu}
  \(\lim_{n\rightarrow\infty} \nu_n = 0\) and for some finite $\Delta > 1$,
  \(\lim_{n\rightarrow\infty}\left(s_{k_n}(\ep_n)\right)^{1
    \over {k_n}}=\Delta.\)
\end{lemma}

The next lemma estimates $L_n(0)$ and $L_n(1)$.

\begin{lemma}\label{lem_lim_L_0_1}
\(\lim_{n\rightarrow\infty}{{L_n(0)} \over {\ln s_{k_n}(\ep_n)}}
\ge {1\over\ln\Delta} \)
and 
\(\lim_{n\rightarrow\infty}{{L_n(1)} \over {\ln s_{k_n}(\ep_n)}}
\ge {\Delta\over\ln\Delta}.\)
\end{lemma}

The next two lemmas estimate
$L_n\left(2\right),\ldots,L_n\left(w-1\right).$ For all integers $n \geq
0$, let \(b_n = \ep_n+C \ln s_{k_n}(\ep_n).\) In light of
Lemma~\ref{lem_s_k_big}, we assume $\ln s_{k_n}(\ep_n) \geq 1$ and thus
$b_n \geq 1$; otherwise we can replace $\{p_n, n \geq 0\}$ with a
subsequence for which these bounds hold.

\begin{lemma}\label{lem_b_n}
  For all $n \geq 0$ and $i \in \{0,1,\ldots,w-1\}$,
  \(b_n^{\left(w-1\right)\left(1+\ep_n\right)^{w-1}} \geq
  s_i(\ep_n).\) 
\end{lemma}

\begin{lemma}\label{lem_L_rest}    
For $j = 2,\ldots,w-1$ and $u=0,\ldots,j-1$,
\( 
\lim_{n\rightarrow\infty} L_n(j) \ge {{\Delta^j}\over{\ln\Delta}}.
\)
\end{lemma}

\subsubsection{Proof of Theorem~\protect{\ref{thm_main}}}
\label{subsub_3}

From (\ref{equ_break_H}) and Lemmas~\ref{lem_lim_L_0_1}
and \ref{lem_L_rest}, we have 
\(
\lim_{n\rightarrow\infty}H(k_n,\vec{s}(\ep_n)) \geq
\frac{1+\Delta+{\cdots}+\Delta^{w-1}}{\ln \Delta}.
\)
By the definition of $r_w$ and the fact $\Delta > 1$ from
Lemma~\ref{lem_nu}, 
\(
\frac{1+\Delta+{\cdots}+\Delta^{w-1}}{\ln \Delta} 
= \frac{\Delta^w-1}{\left(\Delta-1\right)\ln \Delta} \geq C_w.
\)
Combining this with Lemma~\ref{lem_an}, we complete the proof of
Theorem~\ref{thm_main} and thus that of Theorem~\ref{thm_difficult}.

\section{Future research directions} \label{sec_conclusion} 
Our deterministic exploration algorithm is optimal for all $\la$, but our
randomized algorithm is only shown to be optimal for $\la = 1$.  For
general $\la$, it may be possible to obtain better or even optimal
competitive ratios by holding $\la-1$ robots still while one robot moves
towards the origin, as in our deterministic algorithm.  This technique is
not essential in our deterministic algorithm, but there is evidence that it
may be useful in the randomized case.  We further conjecture that there
exists an optimal randomized algorithm in which one robot searches
$w-\la+1$ paths and each of the other robots searches only one of the
remaining paths.

Our exploration algorithms minimize the total distance traveled by all the
robots.  It would be of interest to minimize the total parallel exploration
time.  We conjecture that the optimal competitive ratios of parallel time
are achieved when the paths are partitioned as even as possible.

In the layered graph traversal problem \cite{FFKRRV91,PY91.path,Ramesh95},
one robot searches for a certain goal in a graph.  The robot can shortcut
between paths without going through the origin, and while exploring one
path, it can obtain free information about the other paths. An analog of
our work would be to study how to search a layered graph with multiple
robots.

\section*{Acknowledgments} 
We thank Dan Kleitman for invaluable help with the cyclic assumption of
exploration algorithms and thank Tom Leighton, Ron Rivest, Steve Tate, and
Shang-Hua Teng for many useful discussions.

\appendix

\section{Proof of Lemma~\protect{\ref{lem-sequence}}} \label{app-sequence}

The sequence $\hp$ characterizes a deterministic exploration algorithm
$\AAA(w,1)$, and the smallest competitive ratio of this algorithm is
$1+2{\cdot}\overline{\lim}_{i \rightarrow \infty} H_i$.  If $\AAA(w,1)$
does not have a finite competitive ratio, then Lemma~\ref{lem-sequence}
holds trivially.  Thus, we assume that $\AAA(w,1)$ has a finite competitive
ratio.
\begin{claim}\label{claim_app}
  $\lim_{i \rightarrow \infty} h_i = \infty$.
\end{claim}
\begin{proof}
  Assume for contradiction that the claim does not hold.  Then there exist
  a constant $M < \infty$ and a subsequence $\{ h_{i_k}, k \ge 1 \}$ of
  $\{h_i, i\ge 1\}$ such that $h_{i_k} \le M$ for all $k$.  Therefore,
\begin{eqnarray*}
\overline{\lim}_{i \rightarrow \infty} H_i 
 & \ge  & \overline{\lim}_{k \rightarrow \infty} H_{i_k}\\
 & =    & \overline{\lim}_{k \rightarrow \infty} 
          \frac{h_1 + \cdots + h_{i'_k-1}}{h_{i_k}} \\
 & \ge  & \frac{1}{M}{\cdot}\overline{\lim}_{k \rightarrow \infty} 
          \pa{h_1 + \cdots + h_{i'_k-1}},
\end{eqnarray*}
which equals $\infty$ from the assumption that $\AAA(w,1)$ has a finite
competitive ratio.
\end{proof}

By Claim~\ref{claim_app}, for any $ M < \infty$, $ |\{ h_i \mid h_i \le M
\}|$ is finite.  Hence, we can sort $\{h_i, i \ge 1\}$ into a sorted
sequence $\{ s_i, i \ge 1 \}$. Observe that
\begin{eqnarray}\label{inequ_useful}
  s_1+ \cdots + s_{i} \le h_1+ \cdots + h_{i}, \ \mbox{for all $i \ge 1$.}
\end{eqnarray}
The sequence $\{ s_i, i \ge 1 \}$ is our desired cyclic $w$-sequence, and
its corresponding ratio sequence $\{S_i, i \ge 1\}$ is uniquely defined as
in (\ref{eq-S-def}).
To prove $\overline{\lim}_{i \rightarrow \infty} H_i \ge \overline{\lim}_{i
  \rightarrow \infty} S_i$, it suffices to show that for each $j$ large
enough, there is a $j^*$ such that
\begin{eqnarray}\label{eqn-key}
\mbox{$S_j \le H_{j^*}$ and 
$j^* \rightarrow \infty$ as $j \rightarrow \infty$}.
\end{eqnarray}

For any fixed $j$ that is sufficiently large, we consider  two cases.

{\it Case 1}: There exists some $t \ge j+w-1$ such that $h_t \le s_j$.
Since $t'>t$ is the smallest index with $a_{t'} = a_t$, $t'-1 \ge t \ge
j+w-1$. Hence, by (\ref{inequ_useful}),
\begin{eqnarray}\label{inequ_case1}
S_j = \frac{s_1 + \cdots + s_{j+w-1}}{s_j}
   \le \frac{h_1 + \cdots + h_{j+w-1}}{s_j}
   \le \frac{h_1 + \cdots + h_{t'-1}}{h_t}
    = H_t.
\end{eqnarray}
Let $j^* = t$. Then, (\ref{eqn-key}) follows from
(\ref{inequ_case1}) and the fact $j^* \ge j$.

{\it Case 2}: $h_t > s_j$ for all $t \ge j+w-1$.  Since $\{h_1, \ldots,
h_{j+w-2}\}$ contains all $h_t$ with $h_t \le s_j$, $\{h_1, \ldots,
h_{j+w-2}\}$ contains $\{s_1, \ldots s_j\}$ as a subset.  Thus,
\begin{eqnarray}\label{inequ_set} 
  |\{h_t \mid h_t > s_j \ \mbox{and} \ 1 \le t \le j+w-2 \}| \le \left(j+w-2\right) - j =
  w-2.
\end{eqnarray}
Since $\hp$ is a $w$-sequence, there are $w$ distinct integers $v_1, v_2,
\ldots, v_w$, each appearing infinitely many times in $\{a_i, i \ge 1 \}$.
Since $j$ is sufficiently large, we can assume without loss of generality
that each $v_k$ appears at least once in $\{a_1, \ldots, a_j \}$.  For $k$
with $1 \le k \le w$, let $j(k) \le j+w-2$ be the largest index with
$a_{j(k)} = v_k$.  Among $h_{j(1)},\ldots,h_{j\left(w\right)}$, by
(\ref{inequ_set}), at least two, say, $h_{j\left(k_1\right)}$ and
$h_{j\left(k_2\right)}$, are less than $s_j$.  By the choices of
$j\left(k_1\right)$ and $j\left(k_2\right)$, both $j'\left(k_1\right)$ and
$j'\left(k_2\right)$ are at least $j+w-1$.  Without loss of generality, we
assume $j'\left(k_1\right) > j+w-1$. By (\ref{inequ_useful}),
\begin{eqnarray}\label{inequ_case2}
  \lefteqn{S_j  =  \frac{s_1 + \cdots + s_{j+w-1}}{s_j}}
  \\  & & \ \ \ \ \ \ \le \frac{h_1+\cdots+h_{j+w-1}}{s_j}  \le
  \frac{h_1+\cdots+h_{j'\left(k_1\right)-1}}{h_{l_1}}  =  
  H_{j\left(k_1\right)}. \nonumber
\end{eqnarray}
Now, let $j^* = j\left(k_1\right)$. Since $v_{k_1}$ appears infinitely many
times in $\{a_i, i \ge 1\}$, $j\left(k_1\right) \rightarrow \infty$ as $j
\rightarrow \infty$. Now, (\ref{eqn-key}) follows from
(\ref{inequ_case2}).

Cases 1 and 2 together complete the proof of Lemma~\ref{lem-sequence}.

\section{Proofs of Lemmas \protect{\ref{lem_geo}} through
  \protect{\ref{lem_L_rest}}} \label{app-main}

\subsection{Proof of Lemma \protect{\ref{lem_geo}}}

\begin{eqnarray*}
& {x_1 \over x_0^{1+\ep}} +
{x_2 \over x_1^{1+\ep}} +\cdots+
{x_m \over x_{m-1}^{1+\ep}}  & 
\\
> &
{1 \over {1+\ep}} {x_1 \over x_0^{1+\ep}} + {1 \over \left({1+\ep}\right)^2}
{x_2 \over x_1^{1+\ep}} + \cdots + {1 \over \left({1+\ep}\right)^m} {x_m \over
x_{m-1}^{1+\ep}} & \left(\mbox{because} \ 1+\ep > 1\right)
\\
\geq & 
{1 \over E_\ep\left(m\right)} \left({{x_m^{\left({1 \over {1+\ep}}\right)^m}} \over
x_0}\right)^{E_\ep\left(m\right)} & \left(\mbox{arithmetic mean} \geq \mbox{geometric mean}\right)
\\
\geq & {m \over \left(1+\ep\right)^m} \left({{x_m^{\left({1 \over {1+\ep}}\right)^m}} \over
x_0}\right)^{E_\ep\left(m\right)}. & \left(\mbox{because}\ \left(1+\ep\right)^m \geq 1+m\ep\right)
\end{eqnarray*}

\subsection{Proof of Lemma~\protect{\ref{lem_s_k_big}}}

By the choice of $k_n$ and the monotonicity of
$S_w$, we have $s_{k_n}(\ep_n) \geq s_i(\ep_n)$ for all $i =
0,\ldots,k_n$. Hence,
\begin{equation}\label{equ_L_0}
L_n(0) \geq {{k_n} \over {\left(s_{k_n}(\ep_n)\right)^{\ep_n}}}.
\end{equation}
Then, the lemma follows from the facts that $k_n \rightarrow
\infty$ and $\ep_n\rightarrow 0$ as $n\rightarrow\infty$ and that 
by  (\ref{equ_break_H}) and (\ref{equ_L_0})
\[
C \geq {{-\ep_{n} + {{{k_{n}} \over
      {\left(s_{k_n}(\ep_n)\right)^{\ep_n}}}}}
\over {\ln s_{k_n}(\ep_n)}}.
\]

\subsection{Proof of Lemma~\protect{\ref{lem_nu}}}

First, we have 
\[L_n(1) = \sum_{i=0}^{h_n-1} {{s_{i+1}(\ep_n)}
\over {\left(s_i(\ep_n)\right)^{1+\ep_n}}} + \sum_{i=h_n}^{k_n-1}
{{s_{i+1}(\ep_n)} \over
  {\left(s_i(\ep_n)\right)^{1+\ep_n}}}.
\] 
Noticing  $s_0(\ep_n)=1$ and applying Lemma~\ref{lem_geo} to
the above two summations, we have
\begin{equation}\label{equ_break_L_1}
L_n(1) \geq L'_n(1) + L''_n(1),
\end{equation}
where
\[L'_n(1) = {{h_n} \over {\left(1+\ep_n\right)^{h_n}}}
\left(
\left(s_{h_n}(\ep_n)\right)^{\left({1 \over
{1+\ep_n}}\right)^{h_n}E_{\ep_n}\left(h_n\right)}\right)\]
and
\[L''_n(1) = {{k_n-h_n} \over {\left(1+\ep_n\right)^{k_n-h_n}}}
{\left( {{\left(s_{k_n}(\ep_n)\right)^{\left({1 \over
{1+\ep_n}}\right)^{k_n-h_n}}} \over
{s_{h_n}(\ep_n)}}\right)}^{E_{\ep_n}\left(k_n-h_n\right)}.\]
Now, we can rewrite $L_n''(1)$ as
\[
L''_n(1) = {{k_n-h_n} \over {\left(1+\ep_n\right)^{k_n-h_n}}}
\left({s}_{k_n}(\ep_n)\right)^{\beta''_n},
\] 
where 
\[
\beta''_n = \left(\left({1 \over {1+\ep_n}}\right)^{k_n-h_n}-1+\nu_n\right)
  E_{\ep_n}\left(k_n-h_n\right). 
\] 
We also rewrite $L_n'(1)$ as
\begin{equation} \label{equ:L_1'}
L'_n(1) = {{h_n} \over
{\left(1+\ep_n\right)^{h_n}}}
\left(s_{k_n}(\ep_n)\right)^{\beta'_n},
\end{equation}
where 
\[
\beta'_n = \left(1-\nu_n\right)\left({1
\over{1+\ep_n}}\right)^{h_n}E_{\ep_n}\left(h_n\right).
\]
By Lemma~\ref{lem_s_k_big} and the
fact that by (\ref{equ_break_H}) and (\ref{equ_break_L_1})
\[
C \geq {{-\ep_n +  L'_n(1) + L''_n(1)} \over {\ln s_{k_n}(\ep_n)}},
\] 
we conclude that for some constant $c$,
\begin{equation} \label{equ_L_1_bounded}
0 \leq {{L'_n(1)} \over {\ln s_{k_n}(\ep_n)}} \leq c \ \mbox{and} \ 
0 \leq {{L''_n(1)} \over {\ln s_{k_n}(\ep_n)}} \leq c \ \mbox{for all} \ n.
\end{equation}
Since $1 \leq k_n-h_n \leq w-1$ and $\nu_n \geq 0$, no subsequence
of $\{\beta''_n, n \geq 0\}$ can approach $-\infty$ or
converge to a finite negative number. On the other hand, by
Lemma~\ref{lem_s_k_big} and
(\ref{equ_L_1_bounded}),
no subsequence of 
$\{\beta''_n, n \geq 0\}$ can approach $+\infty$ or converge
to a finite positive number. Thus,
$\lim_{n\rightarrow\infty}\beta''_n = 0$ and consequently,
$\lim_{n\rightarrow\infty} \nu_n = 0.$
By (\ref{equ:L_1'}),  
\[
{{L_n'(1)} \over {\ln s_{k_n}(\ep_n)}}
= {{{h_n}\over{k_n}} \over {\left(1+\ep_n\right)^{h_n}}}
{{\left(\left(s_{k_n}(\ep_n)\right)^{1\over{k_n}}\right)^{k_n\beta'_n}}
\over{\ln \left(s_{k_n}(\ep_n)\right)^{1\over{k_n}}}}.\]
Since $1 \leq k_n - h_n \leq w-1$ and $0 \leq k_n - {1 \over
\sqrt{\ep_n}} \leq w-1$, 
\[\lim_{n\rightarrow\infty} {{{h_n}\over{k_n}} \over
{\left(1+\ep_n\right)^{h_n}}}  =1,
\lim_{n\rightarrow\infty} k_n\beta'_n = 1, 
\ \mbox{and thus}\ 
\lim_{n\rightarrow\infty} {{L'_n(1)} \over {\ln
s_{k_n}(\ep_n)}}=
\lim_{n\rightarrow\infty} 
{\left(s_{k_n}(\ep_n)\right)^{1\over{k_n}}
\over{\ln \left(s_{k_n}(\ep_n)\right)^{1\over{k_n}}}}.
\]
Using Lemma~\ref{lem_s_k_big}, (\ref{equ_L_1_bounded}) and an
argument similar to the proof for $\lim_{n\rightarrow\infty} \nu_n = 0$, we
can show that for some constant $\Delta$
\[\lim_{n\rightarrow\infty}\left(s_{k_n}(\ep_n)\right)^{1 \over
{k_n}}=\Delta > 1. \]

\subsection{Proof of Lemma~\protect{\ref{lem_lim_L_0_1}}}

This lemma follows from Lemma~{\ref{lem_nu}} and the fact that $1 \leq k_n
- h_n \leq w-1$ and $0 \leq k_n - {1\over\sqrt{\ep_n}} \leq w-1$.  The
calculations are similar to those for proving Lemma~{\ref{lem_nu}}.

\subsection{Proof of Lemma~\protect{\ref{lem_b_n}}}

Since $b_n \geq 1$, it suffices to show that for all $i$,
\[
  b_n^{d_i} \geq s_i(\ep_n), 
\]
where $d_i = \sum_{i'=0}^{i-1}\left(1+\ep_n\right)^{i'}$.  We can prove
this inequality by induction on $i$.  The base case follows from the fact
that $s_0(\ep_n) = 1$ and $b_n \ge 1$.  The induction step follows from the
fact that by Fact~\ref{fact_MY} and Lemma~\ref{lem_step1},
\[
C \geq {{-\ep_n + {{s_{i+1}(\ep_n)}\over
      {\left(s_i(\ep_n)\right)^{1+\ep_n}}}} \over {\ln s_{k_n}(\ep_n)}}.
\]

\subsection{Proof of Lemma~\protect{\ref{lem_L_rest}}}
\begin{eqnarray*}
L_n(j) & = & \sum_{u=0}^{j-1} \left(
\sum_{\stackrel{0 \leq i \leq k_n-j}{i \equiv u {\pmod j}}} 
{{s_{i+j}(\ep_n)}\over{\left(s_i(\ep_n)\right)^{\ep_n}}}\right)  
\\
& = & \sum_{u=0}^{j-1}\left(\sum_{i'=0}^{g(j,u)-1}{{s_{u+\left(i'+1\right)j}(\ep_n)}\over
{\left(s_{u+i'j}(\ep_n)\right)^{\ep_n}}}\right) \\
& \geq & \sum_{u=0}^{j-1}L'_n(j,u),
\end{eqnarray*}
where 
\[g(j,u)= \lfloor{{k_n-u}\over j}\rfloor 
\]
and
\[
 L'_n(j,u) = {{g(j,u)} \over
{\left(1+\ep_n\right)^{g(j,u)}}}\left({{\left(s_{h_n}(\ep_n)\right)^{\left({1\over
{1+\ep_n}}\right)^{g(j,u)}}\over{b_n^{\left(w-1\right)\left(1+\ep_n\right)^{w-1}}}}}
\right)^{E_{\ep_n}\left(g(j,u)\right)}.
\]
The term $L'_n(j,u)$ is obtained by applying Lemma~\ref{lem_geo} to the
inner summation in the right-hand side of the above equalities. The
derivation also uses the fact that because $k_n-w+1 \leq u+g(j,u)j \leq
k_n$, 
\[s_{u+g(j,u)j}(\ep_n) \geq s_{h_n}(\ep_n)\] 
and the fact that by Lemma~\ref{lem_b_n}
\[b_n^{\left(w-1\right)\left(1+\ep_n\right)^{w-1}} \geq s_u(\ep_n).\] 
On the other hand, for each $j = 2,\ldots,w-1$ and $u=0,\ldots,j-1$,
\[
\lim_{n\rightarrow\infty} {{L'_n(j,u)}\over{\ln s_{k_n}(\ep_n)}} =
{{\Delta^j}\over{j\ln\Delta}},
\]
which can be verified using Lemma~{\ref{lem_nu}}, the fact $1 \leq k_n -
h_n \leq w-1$ and $0 \leq k_n - {1\over\sqrt{\ep_n}} \leq w-1$, and
calculations similar to those in the proof of Lemma~{\ref{lem_nu}}.

\bibliographystyle{siam}
\bibliography{all}

\end{document}